# Nuclear Equation of State: Picture from Medium Energy Heavy Ion Collisions

Pawel Danielewicz[*]

*Nuclear National Superconducting Cyclotron Laboratory and
Department of Physics and Astronomy,
Michigan State University, East Lansing, MI 48824*
*email:danielewicz@nscl.msu.edu

An equation of state (EOS) is a nontrivial relation between thermodynamic variables characterizing a medium. While the term is used in a singular form in nuclear physics, actually different relations are of interest, such as between pressure $P$, baryon density $\rho$ and temperature $T$, $P(\rho,T)$, between $P$, chemical potential $\mu$ and $T$, $P(\mu,T)$, or between energy per baryon $E$ and $\rho$ and $T$, $E(\rho,T)$, etc. Some of the possible relations are fundamental under certain conditions, meaning that all remaining relations under those conditions may be derived from them, as from $E(\rho)$ at $T=0$.

The nuclear EOS is of interest because it affects the fate of the Universe at times $t>1\mu s$ from the Big Bang and because its features are behind the supernova explosions. Moreover, its features ensure the stability of neutron stars. Through its effects on the evolution of the Universe, on supernovae explosions, and on neutron-star collisions, the EOS affects nucleosynthesis. Moreover, the EOS impacts central reactions of heavy nuclei. Finally, EOS details constrain hadronic interactions and the nonperturbative quantum chromodynamics (QCD).

## Importance of EOS

Different regimes for the strongly interacting are conveniently examined in the $\mu-T$ plane, see Fig. 1. Along the $T=0$ axis, at $\mu\cong930$ MeV, we have the matter inside nuclei. The matter in the interior of neutron stars corresponds to higher chemical potentials, in combinations with low temperatures. The matter in the early Universe evolved along the temperature axis, at low baryon number content, and thus at low $\mu$..

Different regions of the plane are explored at different accelerators. In the early Universe, the matter has crossed the transition between the hadronic matter and quark-gluon plasma. That transition is also likely crossed in the higher-energy accelerators, such as RHIC and SPS. The transition is observed in numerical lattice QCD calculations as a rapid change in energy density in the temperature region of $T_c \cong 170$ MeV [2]. If this transition were of first order at small $\mu$, it could lead to baryon number nonuniformities in the early Universe, impacting nucleosynthesis.

Type II supernova explosions are the source of at least half of the nuclei heavier than iron around us. Only very massive stars, of masses $M > 8M_{Sun}$, explode. Generally, the more massive a star, the shorter it lives, burning faster due to higher density and temperature in its interior. A star starts out burning hydrogen, then helium and successively heavier nuclei; at each stage the products are accumulated. After a given fuel runs out, the gravitation compresses the star core raising temperature and the next fuel ignites with its burning preventing further compression. When the core consists of iron only, the burning stops. It is then up to the electron pressure (such as resisting the compression of solids) to prevent the gravitational collapse of the core. However, the electron pressure fails when the core exceeds the threshold Chandrasekhar mass $M_{th}\cong1.4M_{Sun}$, as easily seen in considering the dependence of core energy on core radius. When the iron core exceeds the threshold, a gravitational collapse of the core starts and progresses until the nuclear densities are reached. The nuclear matter is more incompressible than the electron gas -- what starts as an implosion gets reversed at the nuclear densities into an explosion. From the center of star a shock wave moves out, while at the center a so-called protoneutron star forms at a density of the order of that in nuclei. Inside, as the electron Fermi energy exceeds the proton-neutron mass difference, the process of neutronization takes place, $e^-+p\rightarrow\nu_e+n$. Additionally, thermal neutrinos are copiously produced. In the meantime, the shock moving through the infalling material stalls outside of the protostar and gets, most likely, revived by the neutrinos coming out from the center. Aside from propelling the shock, the neutrinos drive the neutron wind from the center within which copper, nickel, zinc and other elements form. Eventually, the shock reaches the star surface producing a magnificant display in the sky and throwing more than $7M_{Sun}$ of material into space. The properties of nuclear matter, where the collapse reverses and that is the site of neutrino production, are, however, generally not well understood.

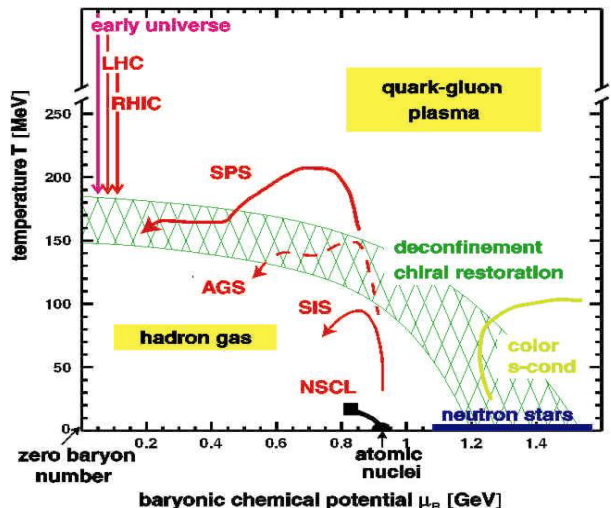

**Fig. 1** Strongly interacting matter in the $\mu-T$ plane.

The protoneutron star eventually turns into a black hole or into a neutron star. Which is the case depends on the properties of nuclear matter [2]. Dependent on those properties are also the characteristics of the forming neutron star. Lattimer and Prakash [3], in particular, find a strong correlation between the radius $R$ of a neutron star and the pressure $p$ at selected densities in neutron matter, when modeling the stars following a host of nuclear EOS, $R\,P^{-1/4} \cong$ const, cf. Fig. 2. Incompressible nuclear matter gives rise to larger neutron stars and yields larger maximal star masses [2,3].

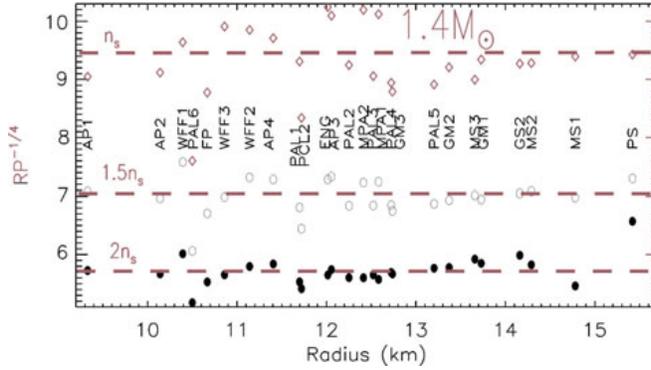

**Fig. 2** Correlation between the radius $R$ of a neutron star and the pressure $P$ at selected densities in neutron matter within the hydrostatic calculations [3] employing different nuclear EOS.

Supernovas are important sites for nucleosynthesis. Possible alternative sites are neutron star mergers, provided the nuclear matter is soft, i.e. pressures in the matter at nuclear densities are relatively low. If the matter is soft, the collisions shed a lot of matter into the space and they shed little matter if the matter is stiff [4], see Fig. 3.

## Elementary Features of Nuclear EOS

Energy per nucleon $E$ in cold nuclear matter can be represented as a sum of the energy of symmetric nuclear matter $E_0$ and a correction $E_1$ associated with the neutron-proton (np) imbalance:

$$E(\rho_n,\rho_p)=E_0(\rho)+E_1(\rho_n,\rho_p) \,. \tag{1}$$

At small relative np asymmetries, on account of the np symmetry of nuclear interactions, the energy $E$ must be quadratic in the asymmetry,

$$E_1 = E - E_0 \approx S(\rho)\left(\frac{\rho_n - \rho_p}{\rho}\right)^2 . \tag{2}$$

Microscopic calculations, such as [5], indicate that Eq. (2) is valid outside of small asymmetries, in fact down to asymmetry of 1 characterizing pure neutron matter, for a wide range of $\rho$. Traditionally, both the energy $E_1$ and the coefficient of proportionality $S$ in (2) are called the symmetry energy. With (2), the energy in neutron matter, characterized by $\rho_n \gg \rho_p$, is given by $E \cong E_0 + S$.

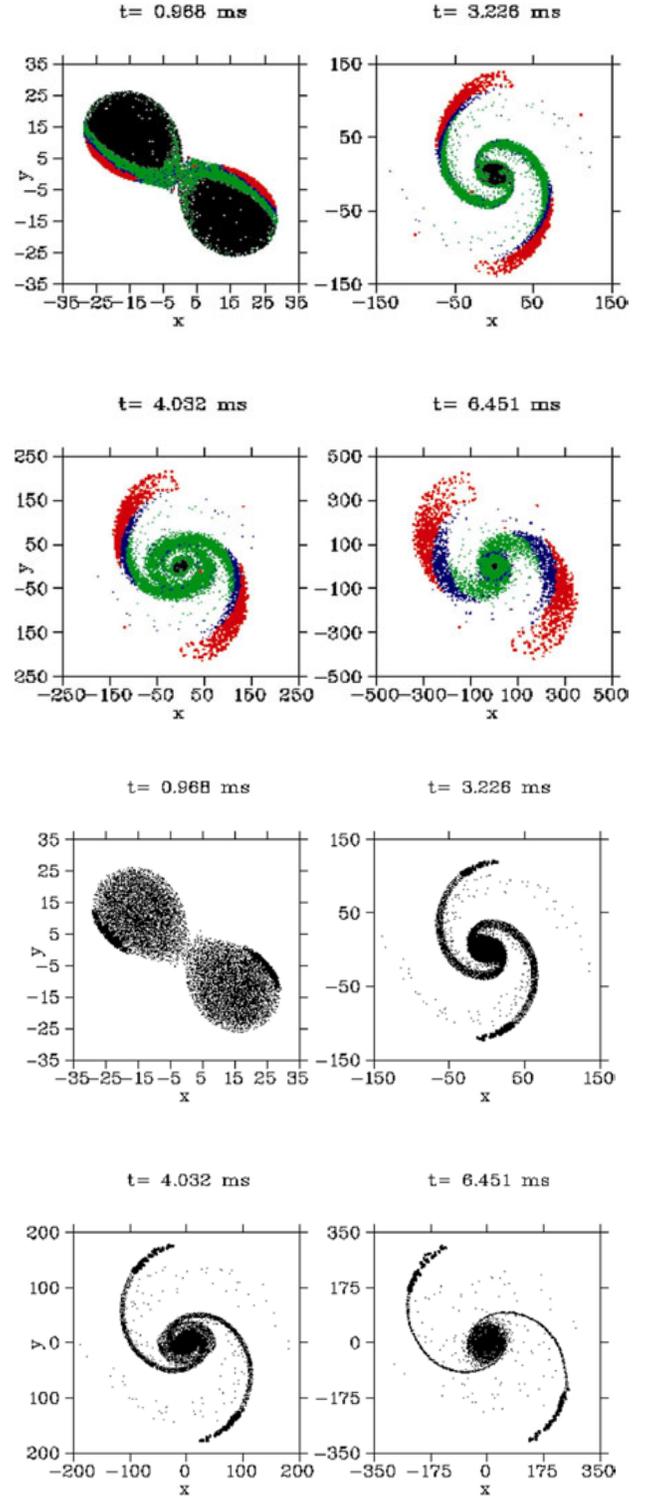

**Fig. 3** Simulations of neutron-star mergers for a soft (upper four panels) and stiff (lower four panels) nuclear EOS [4].

Basic information on nuclear energy stems from electron scattering and from the nuclear binding formula,

$$A\,E(A,Z) = -16\text{MeV}\,A + a_S A^{2/3} + a_C Z^2/A^{1/3}$$
$$+ 21\text{MeV}\,(N-Z)^2/A, \tag{3}$$

where the leading r.h.s. term represents an np symmetric nuclear interior under the sole influence of nuclear interactions. Those sources of information indicate that the energy $E_0$ of symmetric matter minimizes at the normal density $\rho_0$=0.16fm$^{-3}$ with a value of –16MeV relative to nucleon mass, see Fig. 4.

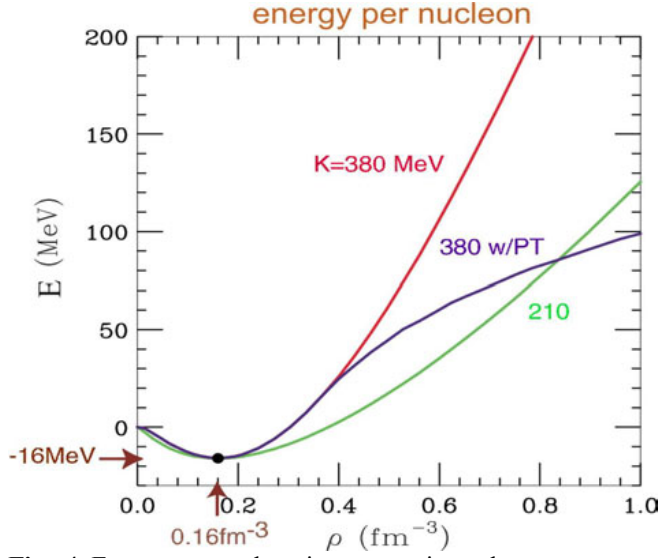

**Fig. 4** Energy per nucleon in symmetric nuclear matter vs density.

Of interest around the minimum is the curvature of energy with respect to density, commonly described in terms of the incompressibility $K$, originally introduced as a curvature of the energy with respect to the nuclear radius $R$:

$$K = 9\rho_0^2 \frac{d^2 E}{d\rho^2} = R^2 \frac{d^2 E}{dR^2}. \quad (4)$$

The incompressibility is studied by inducing collective vibrations of nuclei, changing the density, such as in $\alpha$ scattering [6,7]. Current results indicate that the nuclear incompressibility is in the range [8] $K$=(230-240)MeV, corresponding to a relatively soft EOS around the energy minimum.

Pressure in cold nuclear matter is related to the energy per nucleon with $P = \rho^2 \partial E / \partial \rho$. For densities close to normal, $\rho \sim \rho_0$, because of the minimum in $E_0$, the pressure in neutron matter becomes dominated by the symmetry energy, $P \approx \rho^2 \partial S / \partial \rho$.

**EOS at Supranormal Densities from Flow**

Features of EOS at supranormal densities can be inferred from flow produced in collisions of heavy nuclei at high energies [9]. At low impact parameters, in those collisions, macroscopic regions of high density are formed. The collective flow, which can be quantitatively assessed in collisions, is the particle motion characterized by space-momentum correlations of dynamic origin. The flow can provide information on the pressure generated in the collision.

To see how the flow relates to pressure, we may look at the hydrodynamic Euler equation for the nuclear fluid, an analog of the Newton equation, in a local frame where the collective velocity vanishes, $v$=0:

$$m_N \rho \frac{\partial \vec{v}}{\partial t} = -\vec{\nabla} P. \quad (5)$$

The collective velocity becomes an observable at the end of a reaction. In comparing to the Newton equation, we see that the force potential got replaced by pressure and mass got replaced by mass density. We see from the Euler equation that the collective flow can tell us about the pressure in reaction in relation to density. In establishing the relation, we need to know the spatial size where the pressure gradients develop and this will be determined by the nuclear size. However, we also need to know the time during which the hydrodynamic motion develops and this can represent a problem.

The equilibrium required for hydrodynamics is not quite achieved in reactions and, thus, transport theory is required to establish links between the EOS and observables; the hydrodynamics alone just yields important insights. The reacting system within the transport theory relying on a Boltzmann equation is described in terms of the phase-space distribution functions $f$ for different particles. In particular, the net system energy is a functional of the distributions, $E_{\text{tot}}\{f\}$, and it can be parametrized to yield different EOS in equilibrium limit. The distributions follow a set of Boltzmann equations with single-particle energies representing functional derivatives of the energy, $\varepsilon = \delta E_{\text{tot}}/\delta f$:

$$\frac{\partial f}{\partial t} + \left(\vec{\nabla}_p \varepsilon\right) \cdot \left(\vec{\nabla}_r f\right) - \left(\vec{\nabla}_r \varepsilon\right) \cdot \left(\vec{\nabla}_p f\right) = I, \quad (6)$$

where $I$ is collision integral.

The first observable that one may want to consider to extract the information on EOS is the net radial or transverse collective energy. That energy may reach as much as half of the total kinetic energy in a reaction. Despite its magnitude, the energy is not useful for extracting the information on EOS because of the lack of information on how long the energy develops. Large pressures acting over a short time can produce the same net collective energy as low pressures acting over a long time. This makes apparent the need for a timer in reactions.

The role of the timer in reactions may be taken on by the so-called spectators. The spectator nucleons are those in the periphery of an energetic reaction, weakly affected by the reaction process, proceeding virtually at undisturbed original velocity, see Fig. 5. Participant nucleons, on the other hand, are those closer to the center of the reaction, participating in violent processes, subject to matter compression and expansion in the reaction. As the participant zone expands, the spectators, moving at a prescribed pace, shadow the expansion. If the pressures in the central region are high and the expansion is rapid, the anisotropies generated by the presence of spectators are going to be strong. On the other hand, if the pressures are low and, correspondingly, the expansion of the matter is slow, the shadows left by spectators will not be very pronounced.

anisotropy appears too high at virtually all energies. It should be mentioned that the incompressibilities should be considered here as merely labels for the different utilized EOS. The pressures resulting in the expansion are produced at densities significantly higher than normal and, in fact, changing in the course of the reaction.

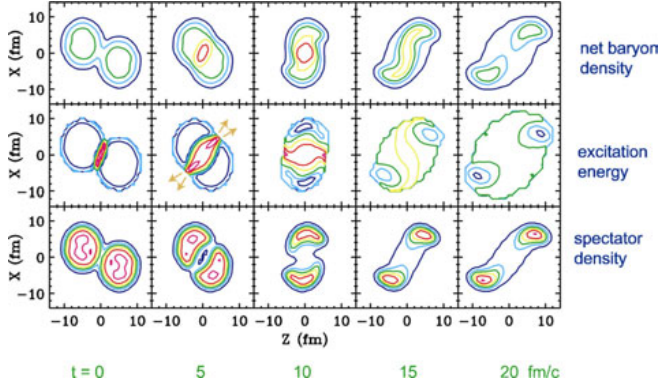

**Fig. 5** Reaction-plane contour plots for different quantities in a $^{124}$Sn+$^{124}$Sn reaction at 800 MeV/nucleon and $b$=6fm, from transport simulations by Shi [10].

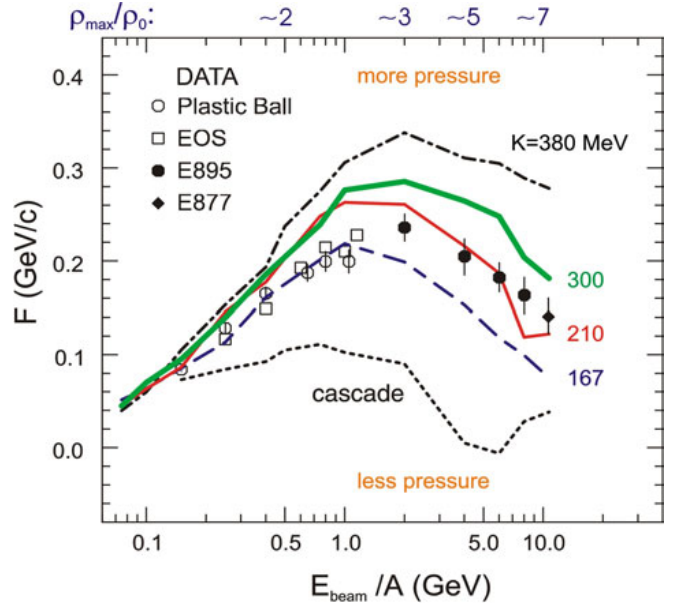

**Fig. 6** Sideward flow excitation function for Au+Au. Data and transport calculations are represented, respectively, by symbols and lines [9].

There are different types of anisotropies in emission that the spectators can produce. Thus, throughout the early stages of a collision, the particles move primarily along the beam axis in the center of mass. However, during the compression stage, the participants get locked within a channel, titled at an angle, between the spectator pieces, cf. Fig. 5. As a consequence, the forward and backward emitted particles acquire an average deflection away from the beam axis, towards the channel direction. Another anisotropy may be observed for particles emitted in the transverse directions with zero longitudinal velocity. The region with compressed matter is open to the vacuum in the direction perpendicular to the reaction plane. However, in the direction within the reaction plane the region is shadowed by the participants. Thus, more particles are expected to be transversally emitted from the participant region, relative to the beam direction, perpendicular to than within the direction plane. The anisotropy should be stronger the faster the expansion of compressed matter.

The different anisotropies have been quantified experimentally over a wide range of bombarding energies. Figure 6 shows the measure of the sideward forward-backward deflection in Au+Au collisions as a function of the beam energy, with symbols representing data. Lines represent simulations assuming different EOS. On top of the figure, typical maximal densities are indicated which are reached at a given bombarding energy. Without interaction contributions to pressure, the simulations labeled cascade produce far too weak anisotropies to be compatible with data. The simulations with EOS characterized by the incompressibility $K$=167MeV yield adequate anisotropy at lower beam energies, but too low at higher energies. On the other hand, with the EOS characterized by $K$=380MeV, the

Figure 7 shows next the anisotropy of emission at midrapidity or zero longitudinal velocity in the c.m., cf. Fig. 8, with symbols representing data and lines representing simulations. Again, we see that without interaction contributions to pressure, simulations cannot reproduce the measurements. The simulations with $K$=167MeV give too little pressure at high energies, and those with $K$=380MeV generally too much. A level of discrepancy is seen between data from different experiments.

We see that no single EOS allows for a simultaneous description of both types of anisotropies at all energies. In particular, the $K$=210MeV EOS is best for the sideward anisotropy, and the $K$=300MeV EOS is the best for the other, so-called elliptic, anisotropy. We can use the discrepancy between the conclusions drawn from the two types of anisotropies as a measure of inaccuracy of the theory and draw broad boundaries on pressure as a function of density from what is common in conclusions based on the two anisotropies. To ensure that the effects of compression dominate in the reaction over other effects, we limit ourselves to densities higher than twice the normal. The boundaries on the pressure are shown in Fig. 9 and they eliminate some of the more extreme models for EOS utilized in nuclear physics, such as the relativistic NL3 model and models assuming a phase transition at relatively low densities, cf. Fig. 9.

Regarding the lower end of densities assessed in Figs. 6-9 and below, comparison of theory to data on kaon production in nuclear collisions below the proton-proton threshold suggests a relatively soft EOS [12], see Fig. 10. The sensitivity is due to the need for multiple collisions needed for production, likely when high densities associated with a soft EOS are reached.

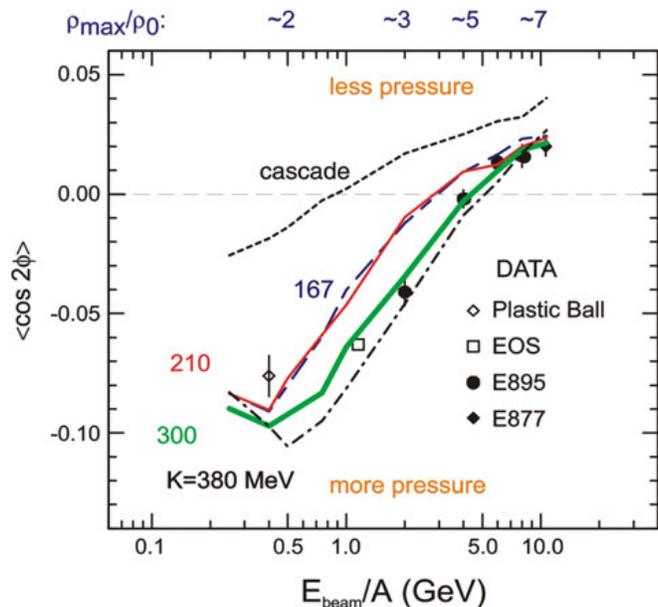

**Fig. 7** Elliptic flow excitation function for Au+Au. Data and transport calculations are respresented, respectively, by symbols and lines [9].

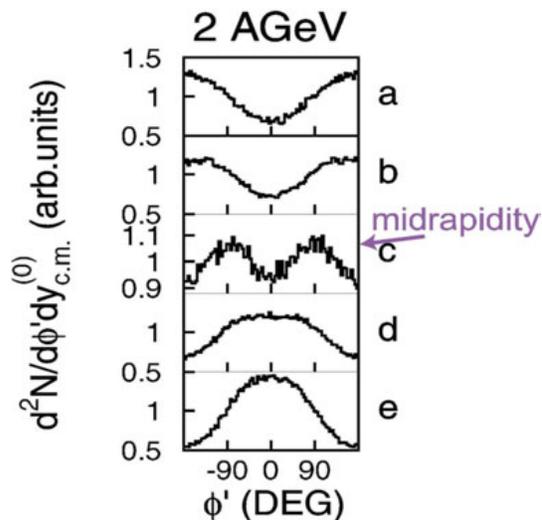

**Fig. 8** Azimuthal distribution of protons from Au+Au collisions at 2GeV/nucleon in different rapidity intervals [11].

When extrapolating to the limit of neutron matter, the symmetry-energy contribution needs to be added to the pressure. The uncertainties associated with the symmetry energy are comparable to those associated with the experimental constraints on the pressure of symmetric matter, and become dominant towards $\rho_0$, see Fig. 11.

## Symmetry Energy from Diffusion in Reactions

The magnitude of symmetry energy at moderately subnormal densities, typifying average densities in nuclei, is constrained by the binding energy formula (3). Different methods have been proposed for constraining the symmetry energy in reactions [14]. One of those has been the impact of symmetry energy on the diffusion [15] of np asymmetry across the reaction zone, in a collision of heavy nuclei of different asymmetry, such as Sn isotopes.

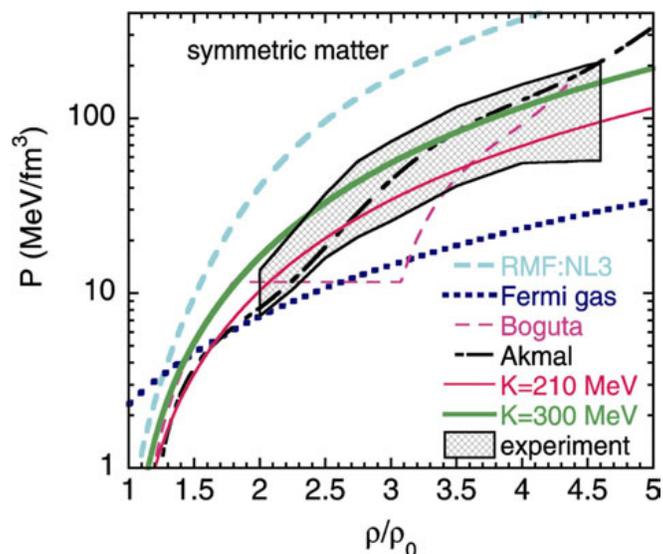

**Fig. 9** Constraints [9] from flow (shaded area) on zero-temperature pressure-density relation of symmetric nuclear matter compared to models for EOS employed in the literature (lines).

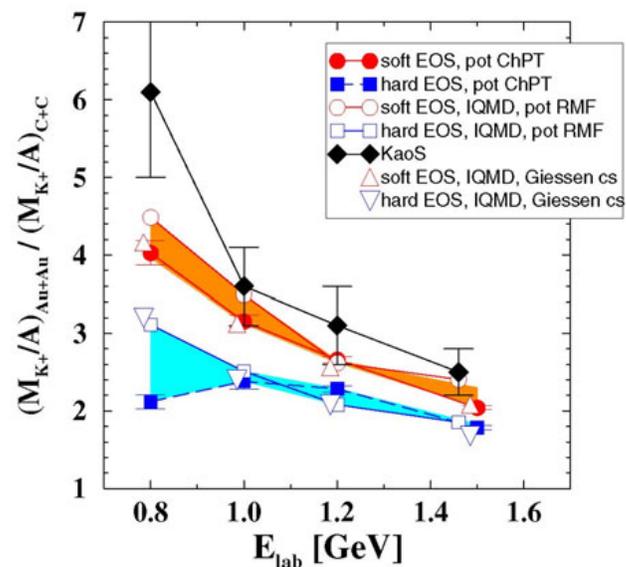

**Fig. 10** Ratio of kaons produced per participant nucleon in Au+Au collisions to kaons per participant nucleon in C+C collisions as a function of beam energy [12]. The filled diamonds represent KaoS data [13] and open symbols represent theory for different EOS.

The irreversible asymmetry flux across the reaction zone is similar to electric current. Driving force for the flux is the difference in gradients for neutrons and protons. The higher the symmetry energy, the stronger the force driving the transport and faster the equilibration of asymmetry across the system. This is illustrated with the results [16] of simulations for reactions of Sn isotopes in Fig. 12.

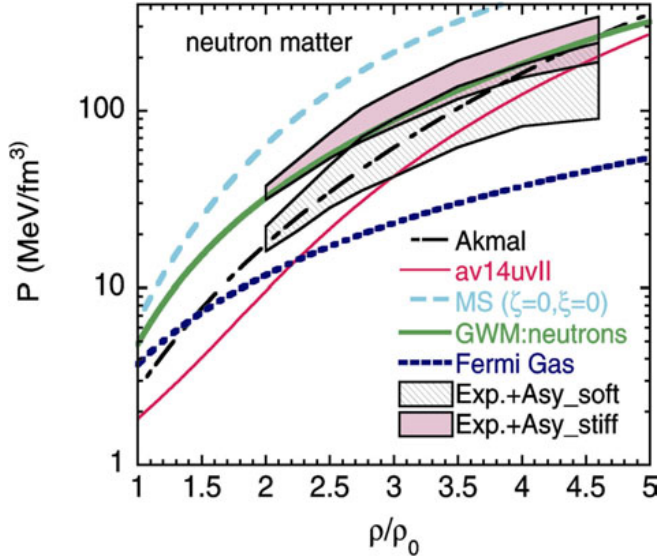

**Fig. 11** Constraints [9] from flow on zero-temperature pressure-density relation of neutron matter, when assuming a symmetry energy with strong (top shaded area) or weak (bottom shaded area) density dependence, compared to models for neutron-matter EOS employed in the literature (lines).

Shown in Fig. 12 is the np asymmetry $\delta=(N-Z)/A$ in the projectile region of Sn systems with different asymmetry, normalized with the projectile-region asymmetries of Sn systems with the same asymmetry. The normalization, yielding the variable $R$ in Fig. 12, aims at illuminating the effects of transport of np asymmetry across a system and suppressing the effects of asymmetry changes in preequilibrium emission and evaporation.

It is found that the symmetry energy with a weaker density dependence produces a faster equilibration for the np asymmetry. This can be understood in terms of the np equilibration taking place at subnormal densities. The data [16] indicate a slow equilibration and, thus, suggest a symmetry energy rapidly changing with density. With regard to the pressure at high nuclear densities, this would imply strong contributions from symmetry energy to the net pressure in Fig. 10. However, the inclusion of more complicated combined momentum and asymmetry dependence in the single-particle energies $\varepsilon$ in (6), implies that the density dependence of symmetry energy could be weaker [17] than suggested in Fig. 11.

**Final Remarks**

Studies of dedicated observables in central reactions of heavy nuclei provide significant constraints on nuclear EOS. The EOS of symmetric matter is found to be rather soft at supranormal densities, but not as soft as when assuming a phase transition at low supranormal densities. In the nearest future, the efforts at EOS determination should aim at providing more stringent constraints on the symmetry energy and at providing ground for the experiments at the forthcoming GSI accelerator, intending to look for the phase transition at finite baryon densities, cf. Fig. 1.

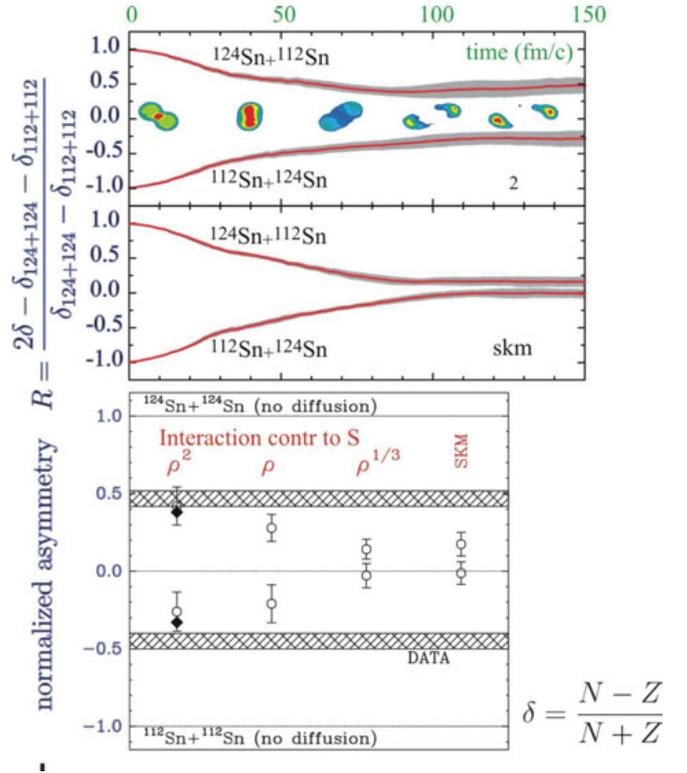

**Fig. 12** Normalized asymmetry $R$ in the projectile region of the $^{124}$Sn+$^{112}$Sn and $^{112}$Sn+$^{124}$Sn reactions at 50MeV/nucleon [16]. The top panel shows the evolution of $R$ in reaction simulations assuming a fast (upper portion) and slow (lower portion) dependence of symmetry energy on $\rho$. The bottom panel compares the results of simulations with different density dependences of symmetry energy (open symbols) to data (filled symbols).